\documentclass[sigconf]{acmart}
\renewcommand\footnotetextcopyrightpermission[1]{}

\usepackage{algorithm}
\usepackage{algpseudocode}
\usepackage{subcaption}

\AtBeginDocument{%
  \providecommand\BibTeX{{%
    \normalfont B\kern-0.5em{\scshape i\kern-0.25em b}\kern-0.8em\TeX}}}

\begin{document}

\title{Guiding Effort Allocation in Open-Source Software Projects Using Bus Factor Analysis}

\author{Aliza Lisan}
\email{alisan@uoregon.edu}
\orcid{0002-1544-1493}
\author{Boyana Norris}
\email{norris@cs.uoregon.edu}
\affiliation{%
  \institution{University of Oregon}
  \streetaddress{1585 E 13th Ave}
  \city{Eugene}
  \state{Oregon}
  \country{USA}
  \postcode{97403}
}

\begin{abstract}
A critical issue faced by open-source software projects is the risk of key personnel leaving the project. This risk is exacerbated in large projects that have been under development for a long time and experienced growth in their development teams. One way to quantify this risk is to measure the concentration of knowledge about the project among its developers. Formally known as the Bus Factor (BF) of a project and defined as “the number of key developers who would need to be incapacitated to make a project unable to proceed” \cite{inproceedings}. Most of the proposed algorithms for BF calculation measure a developer's knowledge of a file based on the number of commits. In this work, we propose using other metrics like lines of code changes (LOCC) and cosine difference of lines of code (change-size-cos) to calculate the BF. We use these metrics for BF calculation for five open-source GitHub projects using the CST algorithm and the RIG algorithm, which is git-blame-based. Moreover, we calculate the BF on project sub-directories that have seen the most active development recently. Lastly, we compare the results of the two algorithms in accuracy, similarity in results, execution time, and trends in BF values over time.
\end{abstract}



\maketitle
\pagestyle{plain}

\section{Introduction} \label{intro}
If we look at our daily use of the Internet, we realize that we are consumed with the use of software and applications. The development and maintenance of all these software projects are based on the knowledge held by its developers. This makes the software development process the most dependent upon its developers. Given that, developers become an important asset for project organizations and open-source project teams. This also makes the rate at which developers leave a software project a critical matter and risk. To mitigate this risk, it is important that project managers or principal developers monitor and quantify the concentration of knowledge of the project among its developers \cite{7961518} An interesting measurement used in this regard is known as the Bus Factor (BF), which is defined as the minimum number of key developers whose departure would make a project unable or difficult to proceed \cite{7961518} A smaller bus factor value would mean that the maximum knowledge of the project is concentrated among them and the project is at a higher risk if some or all of these developers leave the project, company or go for a vacation, etc. A bus factor of one would be the worst-case scenario. Conversely, a high bus factor means a lesser risk is posed to the project in case some of the developers end up leaving.

When it comes to open-source software projects, the bus factor is of much more importance. Most of the people working on these projects are making contributions voluntarily. Volunteering means that the developers or contributors do not have financial benefits associated with these projects. This puts open-source software projects at a higher risk of developer turnover \cite{7961518}. Given the importance of bus factor measurement in open-source software development projects, algorithms have been proposed to calculate it using data gathered from version control systems such as GitHub \cite{inproceedings, 7503718, 10.1145/2884781.2884851}. To the best of our knowledge, most of these algorithms are commit-based, i.e., the algorithms look at the commit data from version control repositories. Moreover, existing work mostly proposes bus factor calculation algorithms and presents tools for the same. Few studies attempt to validate the results of these algorithms. \cite{7961518} is an empirical and comparative study where three Bus Factor algorithms are validated, but they only used the tools and metrics provided by the original authors.

Therefore, in this paper, we look at two algorithms (a) first one was proposed by Cosentino et al. \cite{7961518}, which calculates the bus factor of each file and aggregates it up to branch, directory, or project level; (b) second one proposed by Rigby et al. \cite{10.1145/2884781.2884851}, which uses a git-blame-based approach to calculate the bus factor. The tool based on the first algorithm is publicly available \cite{inproceedings}; however, the authors have only provided the pseudo-code of the second. Using the convention followed in \cite{7961518}, we refer to the first algorithm as CST and the second one as RIG. Among the two algorithms, CST is a commit-based algorithm, and we wanted to use it with other metrics like lines of code changes (LOCC) and cosine difference of lines of code (change-size-cos). For that, we implemented it by understanding the algorithm mentioned in \cite{inproceedings}. RIG algorithm had to be implemented as its code, or any related tool is not publicly available.
We implemented and tested both algorithms on five open-source projects and compared their results. We also got feedback from the principal developers of these projects to validate the results. Lastly, we used the tool provided in \cite{inproceedings} to obtain bus factor results for the selected projects.

To summarize, we seek to answer the following research questions:
\begin{enumerate}
  \item \textbf{How do different metrics such as lines of changed code and cosine difference affect the bus factor computation?}
  \item \textbf{How can the bus factor guide principal developers or managers to allocate effort for hiring or knowledge transfer among the existing team of developers?}
  \item \textbf{How do bus factor algorithms differ in terms of accuracy and performance?}
\end{enumerate}
As such, our contributions can be summarized as follows:
\begin{itemize}
  \item Implemented the CST algorithm (with lines of code changes and cosine difference of lines of code) and the RIG algorithm and compared their results.
  \item Got feedback from the principal developers of the five selected projects to validate our results.
  \item Calculated the BF of the selected projects using the commit-based CST algorithm tool provided in \cite{inproceedings} to compare with our results.
  \item Calculated the bus factor for the selected projects for the last five years to see the trend in BF values.
\end{itemize}

In Section~\ref{lit}, we review the existing literature for bus factor calculation. Section~\ref{study} explains the design and steps involved in our study. We present our results in Section~\ref{results} and conclude the paper in Section~\ref{conclusion}.

\section{Existing Literature} \label{lit}
In this section, first, we describe the two algorithms for measuring bus factors that we will be using as part of our study in this paper. Towards the end, we describe other algorithms that are part of the existing literature but not implemented in our study.

\subsection{CST Algorithm} \label{cst}
The CST algorithm is proposed as part of a tool paper \cite{inproceedings} by Cosentino et al. The algorithm calculates the knowledge of developers on each file and determines the bus factor for each file based on that. Furthermore, aggregation is performed from file to directory, branch, or project level to get developer knowledge. The aggregation is done by simply adding the knowledge on each file and scaling with the number of files in the directory, branch, or project. The authors proposed four metrics to calculate a developer's knowledge of a file. \emph{last change takes it all} assigns 100\% knowledge of a file to the last developer who modified it, and in \emph{multiple changes equally considered}, the knowledge of a developer is calculated by dividing the number of commits made by the developer on a file by the total number of commits ever made on that file. The third metric is \emph{non-consecutive changes}, which considers the developer's non-consecutive commits on a file only and merges the consecutive commits into a single one. The last metric is an extension of the third metric, called the \emph{weighted non-consecutive changes}, and assigns incremental weight to the later commits on the file.

Once the developer's knowledge of the artifacts, i.e., file, directory, branch, or project, has been calculated, the bus factor for each of these artifacts is measured by using two sets of developers. The first set is called the primary developers who have minimum knowledge $X$ of the artifact. The second set is called the secondary developers who at least have some knowledge $Y$ of the artifact, where $Y$ is less than $X$. The threshold $X$ for the selection of primary developers is set to $1/N$ where $N$ is the total number of developers that have made changes to the artifact till date. $Y$ is set of half of $X$ for the selection of secondary developers. The number of developers in the union of both these sets gives us the bus factor for the particular artifact.

Since, CST is a commit-based algorithm, we use the same algorithm with \emph{LOCC} and \emph{change-size-cos}. The details of how we collect and use this data for open-source projects on GitHub are provided in later sections.

\begin{algorithm}
\caption{RIG Algorithm}\label{alg:rig}
\textbf{Data} git-blame data for each file in the project
\newline
\textbf{Result} BFset (BF developers), g (Bus Factor)
\begin{algorithmic}[1]
\For{$g \gets 1 \hspace{0.2cm} to \hspace{0.2cm} 200$}
\For{$i \gets 1 \hspace{0.2cm} to \hspace{0.2cm} 1000$}
\State $BFset \gets $ random sample of g devs;
\State remove-authors (BFset);
\If{ $\emph{abandoned-files()} \ge 50\%$}\\
\hspace{2cm}\Return g, BFset;
\EndIf
\EndFor
\EndFor\\
\Return null, null;
\end{algorithmic}
\end{algorithm}

\subsection{RIG Algorithm} \label{rig}
RIG algorithm was proposed by Rigby et al. \cite{10.1145/2884781.2884851}, in which they adapt the financial risk management measures to a developer turnover context to measure the risk posed to a project from developer turnover. Unlike the CST algorithm, RIG uses a blame-based approach to calculate code ownership of developers. The \emph{blame} feature is implemented in version control systems such that a \texttt{git-blame} command assigns each line in a file to a developer who changed it last. The algorithm's authors claim that this approach allows them to follow code ownership at a finer granularity. As per the algorithm, a line is considered abandoned if it is attributed to a developer who is no longer part of the project, and a file is considered abandoned when 90\% of its files are abandoned. This high threshold makes sure that developers with trivial contributions are excluded.
Algorithm~\ref{alg:rig} shows the pseudo-code for the RIG algorithm. The algorithm starts by varying the size of developers $g$ from 1 to 200 who leave the project, as shown in line 1. Line 3 shows the random sampling of developers of size $g$ who leave the project, and this random sampling is repeated 1,000 times for every value of $g$. We limit these iterations to 10 instead of 1,000 because the algorithm takes a long time on relatively larger projects we chose. Moreover, following \cite{7961518}, we return the lowest value of $g$ that resulted in the abandoned status of more than 50\% of the files, as shown from lines 5-7. The authors in the original paper \cite{10.1145/2884781.2884851} also calculate the likelihood of each group of developers leaving the project between lines 3-4, but we omit it since we are 
A noticeable characteristic of this algorithm is its non-deterministic nature, i.e., different results are produced for each execution of the algorithm because of the random sampling of the developers. Also, the algorithm may not return a valid result, as can be seen from line 10, which happens when no \emph{BFset} results in more than 50\% of \textit{abandon} files.

\begin{table*}
\caption{List of the selected HPC projects with their brief descriptions.}
\label{table:1}
\begin{tabular}{ccccc} 
\toprule
Project & Release Year & Contributors & Language & Description\\ 
\midrule
PETSc\cite{petsc} &1994 &207 &C &Portable, Extensible Toolkit for Scientific Computation.\\ 
Spack\cite{spack} &2014 &1,164 &Python &Multi-platform package manager.\\
Hypre\cite{hypre} &2004 &34 &C &Library of high performance pre-conditioners and solvers.\\
Lammps\cite{lammps} &2016 &224 &C++ &Large-scale Atomic/Molecular Massively Parallel Simulator.\\
NWChem\cite{nwchem} &1994 &44 &Fortran &Open Source High-Performance Computational Chemistry.\\
\bottomrule
\end{tabular}
\end{table*}

\subsection{Other Bus Factor Algorithms} \label{algos}
Another commit-based algorithm was proposed by Avelino et al. and computes the \emph{Degree of Authorship (DOA)} of each file to identify the key developers \cite{7503718}. The DOA value for a file $f$ is initialized when it is created by a certain developer $d$. The DOA of $d$ increases when $d$ makes more commits on $f$ while it decreases if other developers commit to $f$. Lastly, DOA values are normalized for each file with the developer with the highest DOA equal to 1. Developers with a DOA greater than 0.75 are considered the authors of that file. We have not implemented this algorithm as part of our study and chose one commit-based algorithm only i.e., the CST algorithm explained in Section~\ref{cst}.

The first algorithm for the automated calculation of BF from version control systems was proposed by Zazworka et al. \cite{10.1145/1852786.1852805} in which each developer who made changes to a file, regardless of the number of commits, is considered the author of that file. Moreover, to find the BF, they look at each combination of developers ranging from 1 to $N$ (total developers). The BF is the largest combination of developers that a project may lose while the remaining developers still have knowledge of at least a part of the project's files. Since the algorithm looks at each combination, it is shown in \cite{inproceedings-ricca} that the algorithm scales to a maximum of 30 developers only. It is also mentioned in \cite{7961518} that this algorithm did not terminate on projects even after running for more than three days. Hence, we did not include this algorithm in our study since all our projects' total number of developers exceeds 30.

\section{Study Design} \label{study}

In the following subsections, we will discuss in detail the database for GitHub projects that we use in this study, the bus factor algorithms that we implement, and how we validate the bus factor results. For the RIG algorithm, we executed our experimental runs in Google Colaboratory and on a server.

\subsection{Database of GitHub data for HPC projects and other Metrics}
In this work, we used GitHub data from five open-source high-performance computing (HPC) software projects. Table~\ref{table:1} shows some details about these projects. The GitHub data for a number of open-source projects, including these, has been stored, organized and parsed by us in a SQL database to be used for different research projects. The database contains computations for the lines of code changes (\emph{LOCC}) and the cosine difference of code changes (\emph{change-size-cos}), which are used for the bus factor calculation in our study for the CST algorithm.

As mentioned in Section~\ref{lit}, AVL and CST algorithms are commit-based. However, commits are not a great metric to consider to measure code ownership. One reason is the difference in coding styles of different individuals, where some frequently commit after every small change while others prefer to commit once after completing the task. Moreover, a commit-based approach would consider a commit consisting of deletions equal to that of additions regardless of the actual contribution made \cite{10.1145/2884781.2884851}. Also, deletion removes the authorship of a certain piece of code from a developer and, hence, decreases the developer's knowledge. 

With \emph{LOCC}, each file is analyzed based on the number of lines the author has contributed with a commit. However, it will also consider the addition and deletion of blank spaces or comments as meaningful contributions. For the cosine difference of the changes, \texttt{textdistance} Python package is used to examine differences in text. It adds all the words from the changed lines to a word bank, places the added words in a column and deleted words in a row. Then it compares the columns and rows and assigns a numeric value of how much the line was changed. Given that, a change of variable name will be considered as a very small change under the cosine metric but may be a larger change under \emph{LOCC}. Hence, using the cosine difference between the changed lines of code only considers the contributions with actual impact on the files, giving a comprehensive understanding of a developer's contribution.

The authors of the RIG algorithm give a similar reasoning for not using a commit-based approach and instead mention that a blame-based approach allows them to study code ownership at a finer granularity \cite{10.1145/2884781.2884851}.

\begin{algorithm}
\caption{CST Algorithm}\label{alg:cst}
\textbf{Input} cMetric, cstMetric, timeRange (opt.), directory (opt.), branch(es) (opt.)
\newline
\textbf{Result} primaryDev, secondaryDev, busFactor
\begin{algorithmic}[1]
\State $D \gets totalDevelopers$;
\State $primRatio \gets 1/D$;
\State $secondRatio \gets primRatio/2$;
\State Assign dataFrame for sum of cMetric values against each file to $totalPerFileDf$;
\If{$cstMetric == mulChangesEqual$}\
\For{file $f$ in $totalPerFileDf$}
    \State $fileTotal \gets sum(cMetric)$ for $f$;
    \State $devTotal \gets sum(cMetric)$ for $d$ on $f$;
    \State $devKnowledge \gets \frac{devTotal}{fileTotal}$;
\EndFor\
\EndIf\
\State Assign a dataFrame for aggregated $devKnowledge$ to directory/project level to $aggregatedDf$;
\For{$devKnowledge$ in $aggregatedDf$}
    \If{$devKnowledge \ge primRatio$}\
        \State $primaryDev = primaryDev+1$;
    \ElsIf{$devKnowledge \ge secondRatio$}
        \State $secondaryDev = secondaryDev+1$;
    \EndIf\
\EndFor\
\State $busFactor \gets primaryDev$+$secondaryDev$;
\end{algorithmic}
\end{algorithm}

\begin{figure}[ht]
\centering
\includegraphics[width=\linewidth]{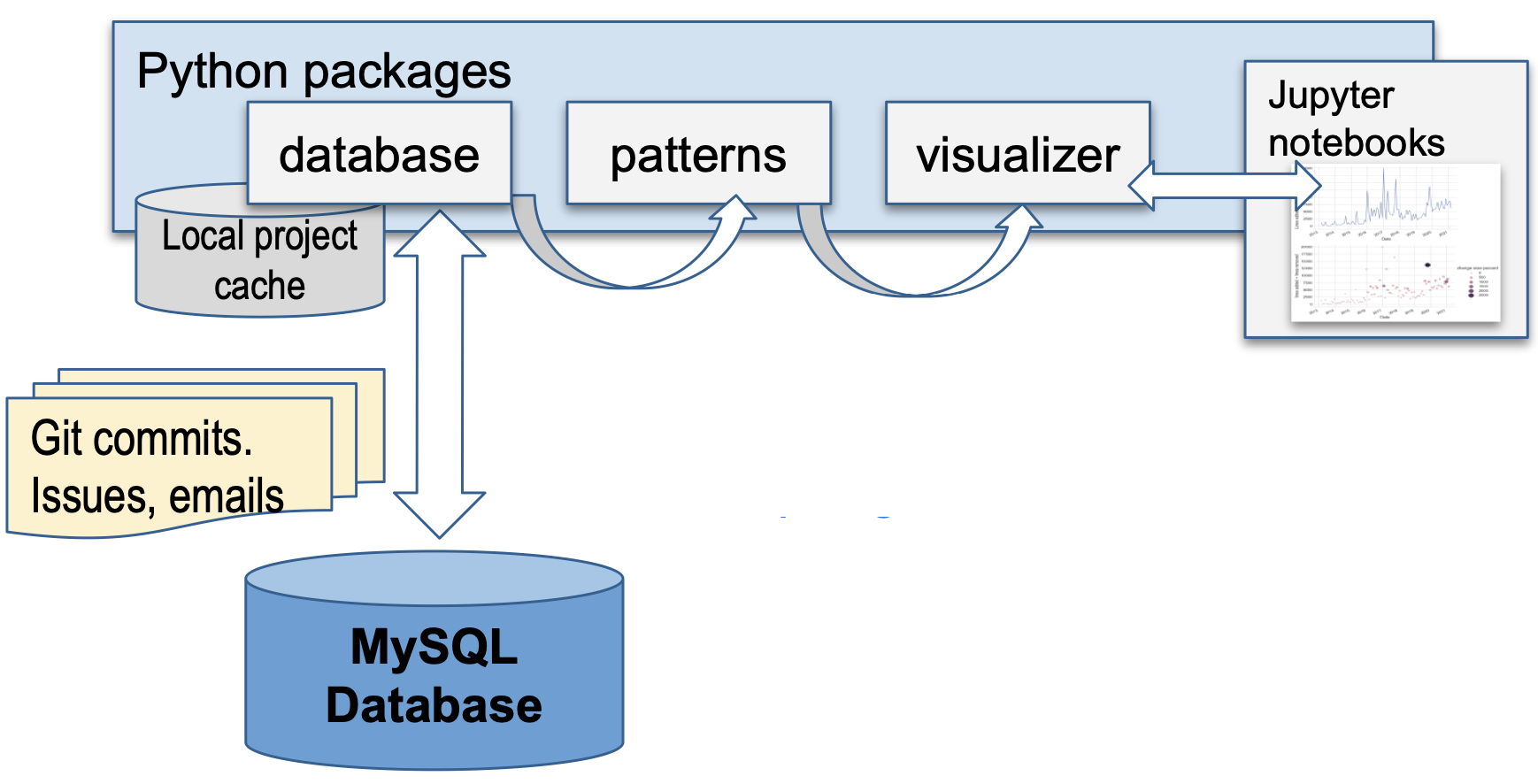}
\caption{GReMCat Software Framework.}
\Description{The software architecture of GReMCat contains three python packages which are database, patterns and visualizer. The database package fetches GitHub data from the database and also caches project related data locally. Patterns package has the implementation of the CST algorithm. The results of the algorithm are displayed by the visualizer package which can be called from Jupyter notebooks.}
\label{fig:1}
\end{figure}

\subsection{Implementation of Bus Factor Algorithms}
After reviewing the literature for the CST algorithm \cite{inproceedings}, we aimed to use this algorithm on two new metrics. Before doing that, we decided to calculate BF on our selected projects using the tools provided by the authors of the CST algorithm. Two tools are needed to calculate the BF of a project, as provided by the authors. The first is called \emph{Gitana}\footnote{https://github.com/valeriocos/Gitana}, which imports all the data from a GitHub repository to an SQL database and then exports it to a JSON file \cite{article}. Second, the web-based bus factor calculating tool\footnote{https://github.com/SOM-Research/busfactor} takes the JSON file as input and calculates BF based on the selected CST metric and thresholds. It is worth noting that exporting data from the database using \emph{Gitana} takes a lot longer than importing the data from GitHub. For instance, running \emph{Gitana} for the smallest project, Hypre took more than 8 hours.

Moreover, we implemented the CST algorithm as part of the Git repository mining and analysis software (GReMCat)\footnote{https://github.com/HPCL/ideas-uo}. The implementation design of the complete software framework can be seen in Figure~\ref{fig:1}. The CST algorithm is implemented in the \texttt{patterns} package and can be used from a Jupiter Notebook\footnote{https://tinyurl.com/CSTnotebook} by calling the \texttt{visualizer} object. The authors of CST do not provide a pseudo-code for their proposed algorithm in their paper \cite{inproceedings}. We implemented it based on the explanation in \cite{inproceedings} and \cite{7961518} and its pseudo-code for our implementation of \emph{multiple changes equally considered} metric is presented in Algorithm~\ref{alg:cst}. Instead of using the number of commits, we used the \emph{LOCC} or \emph{change-size-cos} values against each commit that is pre-computed in our database. The algorithm allows the user to input the branch, directory, time period, CST metric and the proposed data metrics they want to be used for the BF calculation. The time period can be \emph{year-year}, \emph{month-month}, a specific \emph{year} or \emph{month}.

As mentioned in Section~\ref{rig}, the RIG algorithm is not commit-based, and due to the size and continuously changing nature of the git blame data, we do not store it in our database. In fact, we implemented RIG presented in Algorithm~\ref{alg:rig} outside of GReMCat, and executed it on our selected projects in Google Colaboratory and on a server. The authors of the RIG algorithm do not provide a public tool or its code but it was fairly easy to understand and implement from their explanation in \cite{10.1145/2884781.2884851} and also using the pseudo-code provided in \cite{7961518}.

\subsection{Handling External Code and Aliases}
GReMCat already had its implementation of removing external code and libraries from the commits data of the projects in the database based on the names of directories. We used it for the CST algorithm and applied the same methodology to the RIG algorithm. Moreover, GReMCat uses Python fuzzywuzzy package \cite{fuzzy} for identifying multiple different names and emails associated with the same author. We used this implementation for both CST and RIG algorithms to handle aliases.

\begin{table}
\caption{Bus factors provided by the principal developers of the projects and the loss tolerance of each calculated using Equation~\ref{eq}.}
\label{table:2}
\begin{tabular}{ccc} 
 \toprule
 Project &Key developers &Loss tolerance\\ 
 \midrule
 PETSc &7 &4\\ 
 Hypre &17 &3\\
 Lammps &4 &-\\
 NWChem &4 &2\\ 
 \bottomrule
\end{tabular}
\end{table}

\subsection{Validating Bus Factor Results} \label{validate}
In order to validate the Bus Factor results, we reached out to the principal developers of the five projects via email and asked them three questions: (a) Can you estimate the total number of key developers in the project? If your estimate is for a specific time period, please indicate the years. (b) Who do you consider your top project contributors overall or during a specific time period (please indicate which years)? (c) How many key developers could you lose (in a worst-case scenario) and still continue successfully with your project? We received answers from four out of five projects, while some chose not to name the developers. The answers we received are reported in Table~\ref{table:2}, where the key developers correspond to answers to (a) while loss tolerance corresponds to answers to (c).

Additionally, we also used the tool provided by the authors of the CST algorithm to get BF results for the projects. We were able to export GitHub data to JSON files for three of the projects while getting the BF results for only two of them. The tools continued executing for larger projects like PETSc\cite{petsc} and Lammps\cite{lammps} for more than 48 hours without producing results. 

\begin{table}
\caption{Bus Factor values calculated using \emph{change-size-cos} and LOCC in CST, RIG, and the tools by the authors of commits-based CST.}
\label{table:3}
\begin{tabular}{ccccc} 
 \toprule
 Project &CST (cos) &CST (LOCC) &RIG &CST (commits)\\ 
 \midrule
 PETSc &32 &28 &- &-\\ 
 Hypre &12 &10 &29 &5\\
 Lammps &24 &23 &- &-\\
 NWChem &13 &13 &- &7\\
 Spack &211 &172 &- &-\\
 \bottomrule
\end{tabular}
\end{table}

\begin{figure}[ht]
     \centering
     \begin{subfigure}[b]{\linewidth}
         \centering
         \includegraphics[width=\linewidth]{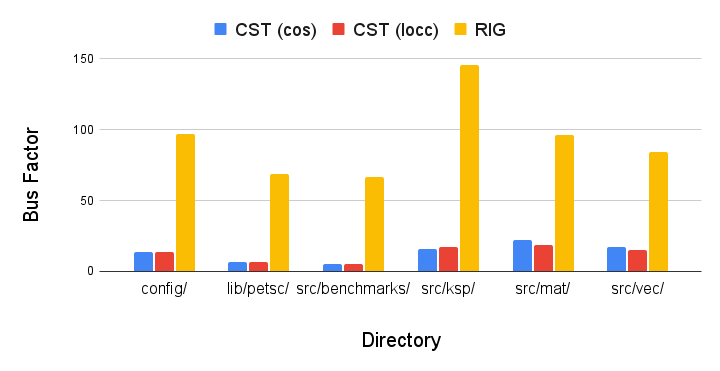}
         \subcaption{PETSc}
         \label{fig:petsc-cst-rig}
     \end{subfigure}
     \hfill
     \begin{subfigure}[b]{\linewidth}
         \centering
         \includegraphics[width=\linewidth]{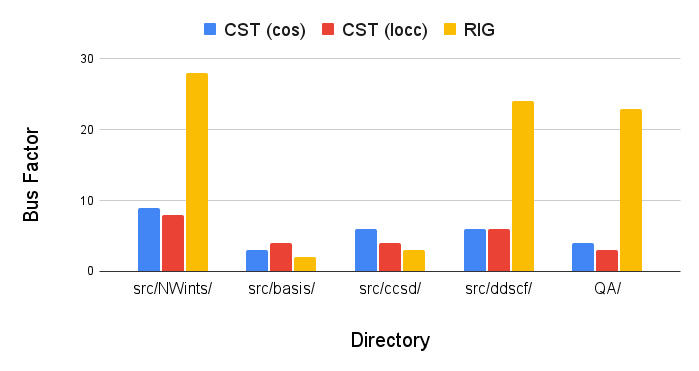}
         \subcaption{NWChem}
         \label{fig:nwchem-cst-rig}
     \end{subfigure}
        \caption{Comparison between LOCC based CST, \emph{change-size-cos} based CST and RIG bus factors for project directories.}
        \Description{The two bar charts for PETSc and NWChem show that the bus factor reported by the RIG algorithm tends to be higher as compared to LOCC and change-size-cos based CST algorithm. Values for the two versions of CST algorithm do not differ much.}
        \label{fig:3}
\end{figure}

\section{Results} \label{results}
In this section, first we compare the project and directory-level BF results of the two algorithms with the feedback received from the principal developers and compare results from different CST metrics. Secondly, we look at the trend in bus factor values for the past five years. Lastly, we look into the performance of RIG algorithm by comparing its directory-level results.

\begin{figure}[ht]
     \centering
     \begin{subfigure}[b]{\linewidth}
         \centering
         \includegraphics[width=\linewidth]{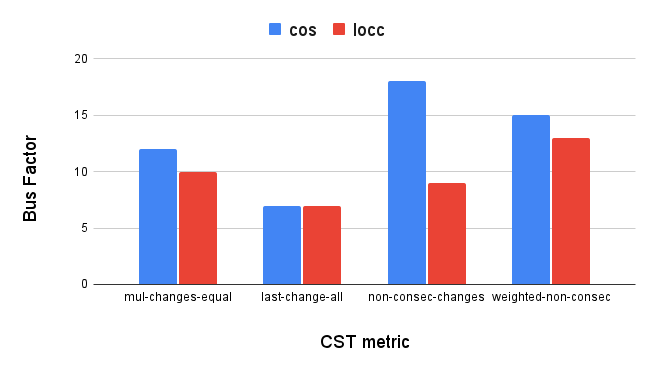}
         \subcaption{Hypre}
         \label{fig:hypre-metrics}
     \end{subfigure}
     \hfill
     \begin{subfigure}[b]{\linewidth}
         \centering
         \includegraphics[width=\linewidth]{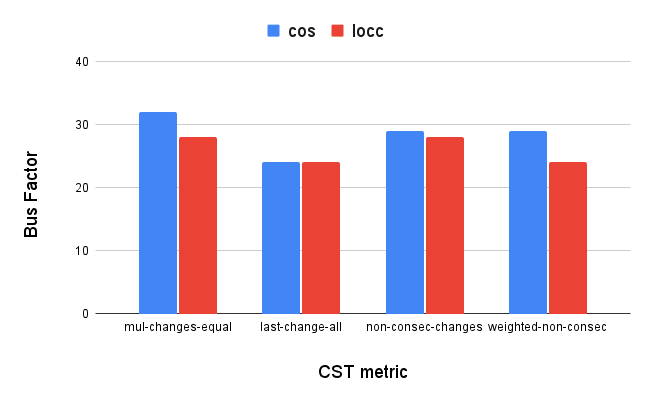}
         \caption{PETSc}
         \label{fig:petsc-metrics}
     \end{subfigure}
        \caption{Comparison between bus factor values for the combinations of the four CST metrics and the two data metrics.}
        \Description{The two bar charts for Hypre and PETSc show that the bus factor values for the four CST metrics and the two proposed metrics do not differ a lot.}
        \label{fig:4}
\end{figure}

\subsection{Comparison for Accuracy of Results} \label{accuracy}
In this subsection, we answer the first of the research questions, which focuses on the impact of different metrics on the bus factor computation. The bus factors calculated by the CST algorithm with our proposed data metrics, the CST tool provided by the authors (commits-based), and our implementation of the RIG algorithm are reported in Table~\ref{table:3}. The values reported are based on the \emph{mul-changes-equal} CST metric. The tool for the CST algorithm took more than 8 hours to produce results for each of the two projects Hypre and NWChem. The calls to the SQL database containing the commits data timed out for the other three projects, given their large sizes. Moreover, the RIG algorithm only returned a bus factor for the Hypre project, which is the smallest in terms of the number of files and subsequently the \texttt{git-blame} data. For projects with many files and authors, the algorithm continued execution for more than 24 hours. Thus, the missing results in the Table~\ref{table:3} highlight the limitations of the RIG algorithm and the tools of the CST algorithm given the large size of the projects.

As mentioned in Section~\ref{validate}, for the accuracy of results and validation of the bus factor values in Table~\ref{table:3}, we got in touch with the principal developers of each of the projects and recorded their responses in Table~\ref{table:2}. By comparing the bus factor values in Table~\ref{table:2} and ~\ref{table:3}, it can be seen that the values given by the developers do not exactly match the values by any of the algorithms. However, the error for the Hypre project, which is calculated as follows:
\begin{equation} \label{eq}
    error = |BF_{algorithm} - BF_{principal\_dev}|
\end{equation}
is the smallest for \emph{change-size-cos} ($error = 5$) and LOCC ($error = 7$) based CST algorithm implemented as part of this study. $Error = 12$ for the commit-based CST \cite{inproceedings} and the RIG algorithm \cite{10.1145/2884781.2884851}. Lastly, we looked at the results from the commit-based CST algorithm and observed that the results were not in sync with the cut-offs mentioned in \cite{inproceedings} for primary and secondary developers, hence, the high error value.

\begin{figure}[ht]
     \centering
     \begin{subfigure}[b]{\linewidth}
         \centering
         \includegraphics[width=\linewidth]{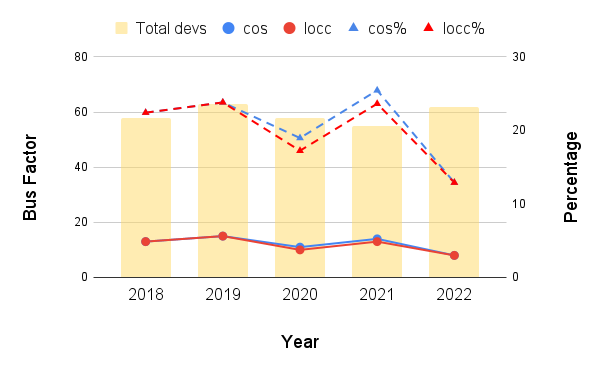}
         \subcaption{PETSc}
         \label{fig:petsc-trend}
     \end{subfigure}
     \hfill
     \begin{subfigure}[b]{\linewidth}
         \centering
         \includegraphics[width=\linewidth]{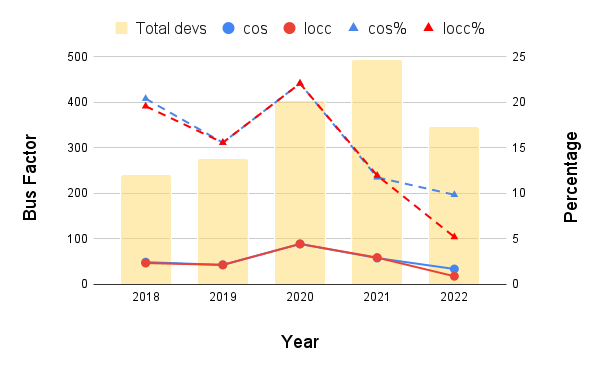}
         \subcaption{Spack}
         \label{fig:spack-trend}
     \end{subfigure}
        \caption{Trend in project level bus factors for last five years for LOCC and \emph{change-size-cos} based CST. The yellow bars represent the total number of developers. The dotted lines represent the percentage of BF developers from the total corresponding to the right y-axis.}
        \Description{The two line charts for PETSc and Spack have the years from 2018 to 2022 on the x-axis. There are two solid and two dotted lines of blue and red color each, corresponding to change-size-cos and LOCC based CST algorithm.}
        \label{fig:5}
\end{figure}

Along with comparing the bus factor values, we also looked at their accuracy in identifying the developers who are part of the bus factor result by each algorithm. Only the principal developers of the Lammps and NWChem project shared the names of the developers constituting the bus factor. For privacy reasons, we won't be sharing the names in this paper. However, from our comparative study, we can state that the developers identified by the principal developers for both Lammps and NWChem are at the top of the sorted list for \emph{change-size-cos} and LOCC-based CST algorithm. The top developer identified by the tool in \cite{inproceedings} for NWChem is not even part of the developers named by the NWChem's principal developer. This shows that the threshold for primary and secondary developers in \cite{inproceedings} is not the best case and also that \emph{change-size-cos} and LOCC-based CST algorithm is more accurate. RIG performed the worst by identifying only 5 out of 12 (\emph{change-size-cos}) or 5 out of 10 (LOCC) developers correctly for the Hypre project. It is important to note that the five projects chosen for this study are large in terms of files, number of developers, commits, and \texttt{git-blame} data. This highlighted the limitation of \cite{article}, \cite{gitana}, and \cite{bus-factor} when it comes to large GitHub projects, as results were not returned for the three larger projects.

In Figure~\ref{fig:3}, we look into the bus factors for the most recently updated directories of PETSc and NWChem projects. We compare the bus factor values of the \emph{change-size-cos} and LOCC-based CST algorithm with the RIG algorithm. It can be clearly seen that the results from the RIG algorithm are significantly different and higher than not only the CST algorithm but also the data provided by the principal developers of the projects. Hence, we found the RIG algorithm to be the worst performer.

Moreover, we present comparative plots for the four CST algorithm metrics explained in Section~\ref{cst} and also the two data metrics \emph{change-size-cos} and LOCC in Figure~\ref{fig:4} for Hypre and PETSc. It can be seen from the graphs that there is not a significant variation in the results for each combination. Given that information, we focused our study on the \emph{mul-changes-equal} metric. However, as mentioned in \cite{inproceedings}, a CST metric can be chosen as per the requirements and processes followed by an organization for version control.

\subsection{Trend in bus factors over time}
In this subsection, we seek the answer to the second research question, i.e., the role of the bus factor in guiding effort allocation towards hiring and knowledge transfer. The intention is to study the applicability of the bus factor and the guidance these results provide to principal developers and managers. Intuitively, the bus factor of a project or directory will indicate the concentration of knowledge within a certain number of people. If that number is small, then principal developers or project managers can work on effort allocation for hiring new people to work on that project. Another thing that can be done is to ensure knowledge transfer to other developers within the organization. For our study, we looked at the trend in bus factor over the past five years for the selected projects. Figure~\ref{fig:5} shows the trend in bus factor since 2018 for PETSc and Spack. The yellow bars represent the total number of developers; the left y-axis, along with solid lines, represents the bus factor value, while the right y-axis and the dashed lines are for the percentage of developers that are part of the bus factor. For both projects, we can see that the developers with higher knowledge of the projects are decreasing in later years, reaching the minimum value in the year 2022. Knowledge of this trend can help the principal developers or managers hire new people or initiate the knowledge transfer process to other existing developers for these projects.

\begin{figure}[ht]
     \centering
     \begin{subfigure}[b]{\linewidth}
         \centering
         \includegraphics[width=\linewidth]{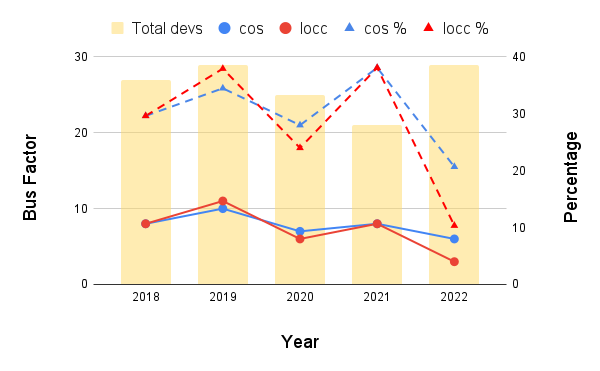}
         \subcaption{PETSc: \texttt{src/ksp/}}
         \label{fig:ksp-trend}
     \end{subfigure}
     \hfill
     \begin{subfigure}[b]{\linewidth}
         \centering
         \includegraphics[width=\linewidth]{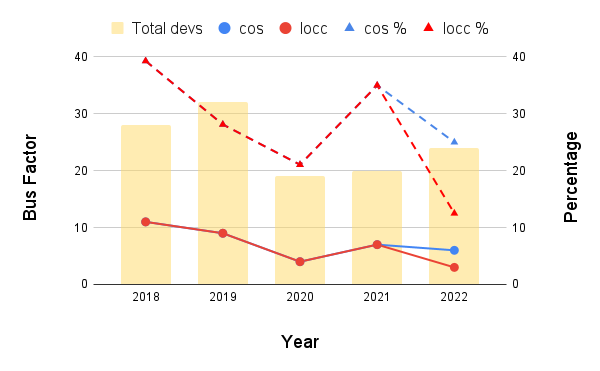}
         \subcaption{PETSc: \texttt{src/mat/}}
         \label{fig:mat-trend}
     \end{subfigure}
     \hfill
     \begin{subfigure}[b]{\linewidth}
         \centering
         \includegraphics[width=\linewidth]{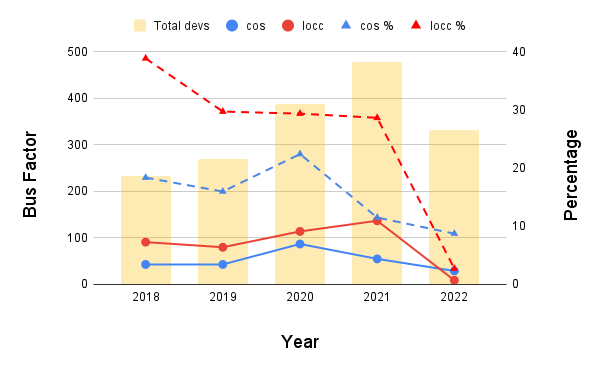}
         \subcaption{Spack: \texttt{var/spack/}}
         \label{fig:vec-trend}
     \end{subfigure}
        \caption{Trend in directory level bus factors for last five years for LOCC and \emph{change-size-cos} based CST. The yellow bars represent the total number of developers. The dotted lines represent the percentage of BF developers from the total corresponding to the right y-axis.}
        \Description{The three line charts for PETSc directories have the years from 2018 to 2022 on the x-axis. There are two solid and two dotted lines of blue and red color each, corresponding to change-size-cos and LOCC based CST algorithm.}
        \label{fig:6}
\end{figure}

Sometimes, looking at the trend in bus factor for the whole project does not provide the best view of the knowledge concentration and the risk of the project in terms of employee turnover. Given that, looking at individual directories or important parts of the projects is more helpful. Figure~\ref{fig:6} shows the trend in bus factor values for major directories of PETSc and Spack. With this narrowed-down view, a more informed decision can be taken for the important parts of the projects. Figure~\ref{fig:6} (c) particularly shows a steep drop in the bus factor value for \texttt{var/spack/} directory of the project, which cannot be picked up from the holistic view shown in Figure~\ref{fig:5} (b).

\subsection{Performance comparison of BF Algorithms}
This subsection focuses on answering the research question about the accuracy of results and the performance of the bus factor algorithms. For that, we wanted to look at each algorithm to compare and contrast results. Most of this question is answered for the CST algorithm in Section~\ref{accuracy}, where we compared the results from the CST algorithm with RIG using Figure~\ref{fig:3}, so here we focused on the RIG algorithm and its non-deterministic nature.

\begin{figure}[ht]
     \centering
     \begin{subfigure}[b]{\linewidth}
         \centering
         \includegraphics[width=\linewidth]{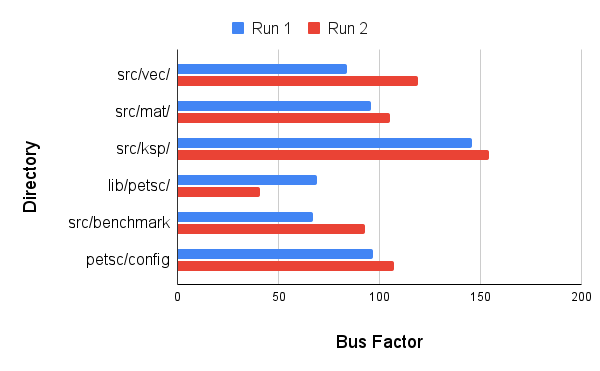}
         \subcaption{PETSc}
         \label{fig:petsc-rig}
     \end{subfigure}
     \hfill
     \begin{subfigure}[b]{\linewidth}
         \centering
         \includegraphics[width=\linewidth]{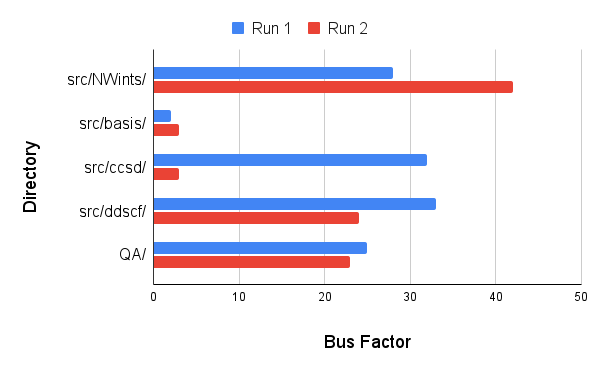}
         \subcaption{NWChem}
         \label{fig:nwchem-rig}
     \end{subfigure}
        \caption{Comparison between bus factors for the two RIG executions.}
        \Description{This figure contains bar charts for PETSc and NWChem directories. The x-axis is the bus factor value while the y-axis has directories. We reported results for six PETSc directories and five NWChem directories.}
        \label{fig:7}
\end{figure}

As mentioned in Section~\ref{rig}, the RIG algorithm is non-deterministic, i.e., it does not produce the same results for different executions. We collected data from two runs for different directories of the five projects to see the variation in results by the RIG algorithm. In Figure~\ref{fig:7}, we show results for two of the projects, PETSc and NWChem. It can be seen that none of the executions resulted in the same result for any of the directories. The difference is significant for some, including \texttt{petsc/src/vec} in (a) and  \texttt{nwchem/src/ccd} in (b). We also looked at the developer names resulting from the algorithm for each execution and did not find a common pattern of similarity among results, which further confirms the random nature of the RIG algorithm.

The non-deterministic and random nature of the RIG algorithm automatically makes the CST algorithm a better choice. This is further corroborated by the responses of the principal developers, as shown in Table~\ref{table:3} and the lower errors for the CST algorithm. Moreover, Rigby et al. themselves claim in \cite{10.1145/2884781.2884851} that bus factor scenarios computed using loss percentages are unrealistic.

\section{Future Work} \label{future}
There are many future dimensions to this work. One way to extend this work is to look at the impact of different threshold values for the algorithms with the proposed data metrics. Since authors of both CST and RIG algorithm do not give reasoning behind the choice of all the chosen threshold and cutoff values, this can be an interesting study. The results can help in making a more informed choice behind these values with reasoning provided.
Moreover, using a more data-driven approach for the validation of results with trained machine learning models can help in eliminating the threats to correctness of the validation process. The large amount of mined GitHub data can be used for training purposes.

\section{Conclusion} \label{conclusion}
It is common for open-source projects to face the risk of key developers leaving the project. This risk is even higher with long-term and large projects. To judge the risk posed to the projects or the severity of it, the concentration of knowledge among its developers can be measured. This measurement is known as the bus factor of the project and there are several algorithms proposed for its calculation. Most of these algorithms use the commits data from version control systems. We proposed the use of two other metrics i.e. lines of code changes (LOCC) and cosine difference of lines of code (\emph{change-size-cos}) and tested them with the CST algorithm. We also used a \texttt{git-blame} based RIG algorithm on our use-case of five High Performance Computing projects on GitHub. Along with complete projects, we also looked at the bus factor values of the major directories in each. We did a comparative study for the accuracy, similarity in results and trend in BF values on the two algorithms and also the metrics proposed by us. We validated our results from the principal developers of the selected projects which showed LOCC and \emph{change-size-cos} to be more accurate than commits. Our implementation of the CST algorithm is scalable when compared to the online available tools by Cosentino at el. Lastly, we demonstrated with examples how looking at the trend in bus factor values over time can guide principal developers in effort allocation.

\bibliographystyle{unsrt}
\bibliography{main}

\end{document}